# Evaluation of Large Language Models for Numeric Anomaly Detection in Power Systems

Yichen Liu, Hongyu Wu
Mike Wiegers Department of Electrical and Computer Engineering
Kansas State University
Manhattan, United States

Bo Liu
School of Engineering and Applied Sciences
Washington State University
Richland, United States

***Abstract*—Large language models (LLMs) have gained increasing attention in power grids for their general-purpose capabilities. Meanwhile, anomaly detection (AD) remains critical for grid resilience, requiring accurate and interpretable decisions based on multivariate telemetry. Yet the performance of LLMs on large-scale numeric data for AD remains largely unexplored. This paper presents a comprehensive evaluation of LLMs for numeric AD in power systems. We use GPT-OSS-20B as a representative model and evaluate it on the IEEE 14-bus system. A standardized prompt framework is applied across zero-shot, few-shot, in-context learning, low rank adaptation (LoRA), fine-tuning, and a hybrid LLM-traditional approach. We adopt a rule-aware design based on the three-sigma criterion, and report detection performance and rationale quality. This study lays the groundwork for further investigation into the limitations and capabilities of LLM-based AD and its integration with classical detectors in cyber-physical power grid applications.**

## I. INTRODUCTION

Large language models (LLMs) have demonstrated impressive versatility across a range of domains, including natural language processing [1], code generation [2], and decision support [3]. Their capability to generalize across tasks with minimal supervision has prompted growing interest in their application to electric power grids.

Existing studies have primarily explored the deployment of LLMs in power grids from two complementary perspectives: ***i)*** the exploration of task-level improvements enabled by LLMs [4] and ***ii)*** the technical integration of LLMs into system workflows [3]. For the former, Majumder *et al.* conducted extensive experiments ranging from prompt engineering to fine-tuning and even integrating LLMs into simulation tools or orchestration frameworks for system analysis. For the latter, Madani *et al.* organized the implementation directions of LLMs across multiple power system domains, including operation and management, market and trading, planning, education, security and compliance. These studies collectively demonstrate the promising potential of LLMs in the electric power sector.

Nevertheless, these studies predominantly adopt a high-level perspective and specific downstream tasks remain insufficiently explored, particularly those involving structured numeric telemetry that require precise reasoning beyond language-based interaction. One such task is numeric anomaly detection (AD), which plays a critical role in maintaining power system stability and resilience by identifying abnormal patterns in multivariate sensor data such as voltages and power injections.

While most existing research integrates LLMs into broader system frameworks using log or text-based inputs, their application to numeric anomaly detection remains limited [5]. This gap is partly due to the tokenization and decoding challenges posed by high-dimensional numeric telemetry [6]. Nonetheless, LLMs' strong reasoning and arithmetic abilities suggest potential for such tasks [4], but their performance boundaries in this context are not yet well understood.

This paper aims to fill this gap by presenting a comprehensive benchmark study that evaluates the performance of LLMs on numeric AD in power systems. We propose a standardized, interpretable prompt structure that incorporates step-by-step, rule-based reasoning to support consistent evaluation across various experimental configurations. The benchmark spans five settings: ***i)*** zero-shot prompting, ***ii)*** few-shot prompting with labeled exemplars, ***iii)*** in-context learning (ICL) with extended examples, ***iv)*** supervised fine-tuning using parameter-efficient low rank adaptation (LoRA), and ***v)*** a hybrid approach that combines LLM reasoning with traditional deep learning (DL) detector. The main contributions of this work are as follows:

- To the best of our knowledge, this study is the first to investigate the application of LLMs for numeric AD in power systems. Our work bridges the gap between LLM reasoning and structured measurements.
- A comprehensive evaluation framework is developed, consisting of ***i)*** multiple prompt-engineering strategies for zero-shot, few-shot, and ICL, ***ii)*** parameter-efficient fine-tuning using LoRA, and ***iii)*** a hybrid configuration that integrates LLM-driven feature selection with a standard deep-learning AD approach.
- A reproducible benchmarking platform is developed using GPT-OSS-20B on the IEEE 14-bus system. The platform enables systematic evaluation of detection performance using classification metrics, e.g., accuracy, precision, recall, and F1-score, and provides a unified environment for assessing LLM behavior on high-dimensional numerical telemetry. This foundation further supports future integration of LLM-based feature extraction with traditional AD methods in power systems.

The remainder of this paper is organized as follows: Section II provides a detailed introduction to LLMs. Section III discusses prompt design and introduces a series of evaluation methods. Section IV presents case studies on the IEEE 14-bus system. Finally, Section V concludes the paper.

## II. Foundations of Large Language Models

LLMs are DL models trained on large-scale text corpora and can both understand and generate human-like language [7]. Compared to conventional pretrained language models (PLMs), they use more parameters and data to handle longer inputs and capture richer context [8]. Most LLMs are built on the Transformer architecture, which underpins their scalability and broad applicability.

### A. Transformer Architecture

The transformer architecture, first introduced by Vaswani *et al.* in 2017, has become the foundation of modern LLMs due to its scalability and strong performance across a wide range of tasks [9]. The overall structure is illustrated in Fig. 1.

*Encoder–decoder:* The Transformer adopts an encoder–decoder structure with six identical layers each. Encoder layers include multi-head self-attention and feedforward networks, while decoder layers add cross-attention over encoder outputs. All sub-layers use residual connections, layer normalization, and share a consistent output dimension of 512.

*Self-attention:* Self-attention, as shown in Fig. 2 (a), maps a query ($Q$) and a set of key-value ($K-V$) pairs to an output, with all elements represented as vectors. The input for self-attention consists of $Q$, $K$ and $V$ of dimension $d_k$, where $d_k$ are the dimension of $Q$, $K$, and $V$. $Q$, $K$, and $V$ are computed by the corresponding weight matrices $W$ by input $x$:

$$Q = x * W(q) + B_q \quad (1)$$

$$K = x * W(k) + B_k \quad (2)$$

$$V = x * W(v) + B_v \quad (3)$$

To compute self-attention, each query vector $Q$ is multiplied by the transpose of the key matrix $K^T$, scaled by $\sqrt{d_k}$ to avoid large values, and passed through a softmax function to produce attention weights. These weights are then applied to the value matrix $V$. In practice, queries, keys, and values are processed in batches as matrices $Q$, $K$, and $V$, with attention computed as:

$$single\ self-attention_i(Q_i, K_i, V_i) = softmax\left(\frac{Q_i K_i^T}{\sqrt{d_k}}\right) V_i \quad (4)$$

The final self-attention output is the average over all individual attention computations:

$$self-attention\ score = \frac{\sum single\ self-attention_i}{Number\ of\ i} \quad (5)$$

*Multi-head attention:* In Fig. 2 (b), multi-head attention projects the input $Q$, $K$, and $V$ into $h$ lower-dimensional subspaces using learned linear mappings. Each head computes attention independently over dimensions $d_k$ and $d_v$, and the results are concatenated and projected back to the original model dimension $d_{\text{model}}$. The multi-head attention is defined as:

$$MultiHead(Q, K, V) = Concat(head_1, \dots, head_H) W^O \quad (6)$$

where each attention head is computed as:

$$head_i = Attention(QW_i^Q, QW_i^K, KW_i^V) \quad (7)$$

Here, $W_i^Q, W_i^K, W_i^V \in \mathbb{R}^{d_{\text{model}} \times d_{Q,K,V}}$ are learned projection matrices specific to the $i$-th head, and $W^O \in \mathbb{R}^{hd_v \times d_{\text{model}}}$ is the final output projection. This design allows the model to attend to information from multiple representation subspaces simultaneously.

*Feedforward Layer*: The feedforward layer applies a fully connected network to each position independently, using a ReLU followed by a linear activation:

$$FeedForward(X) = max(0, xW_1 + b_1) W_2 + b_2 \quad (8)$$

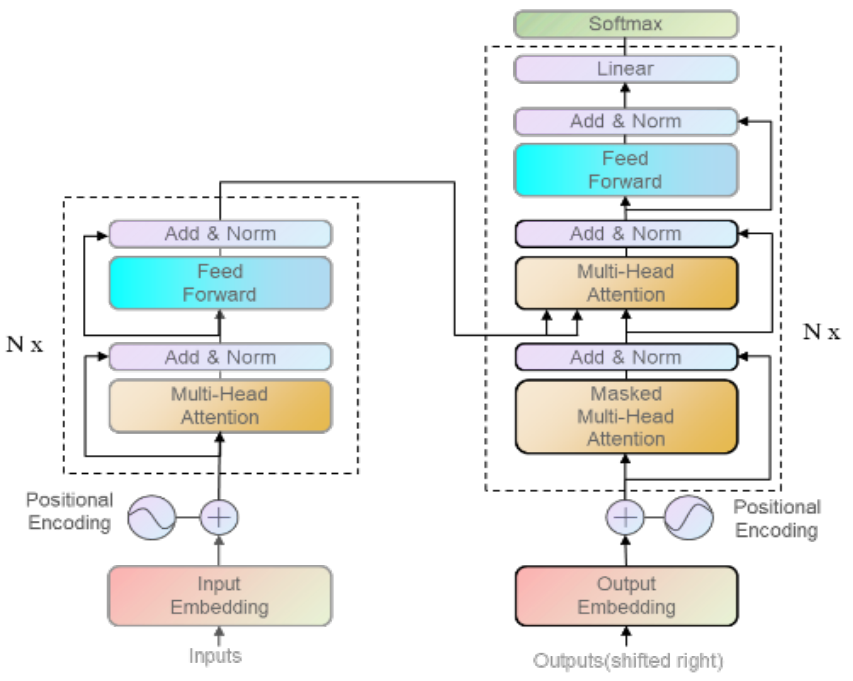

Fig. 1. The classic transformer module

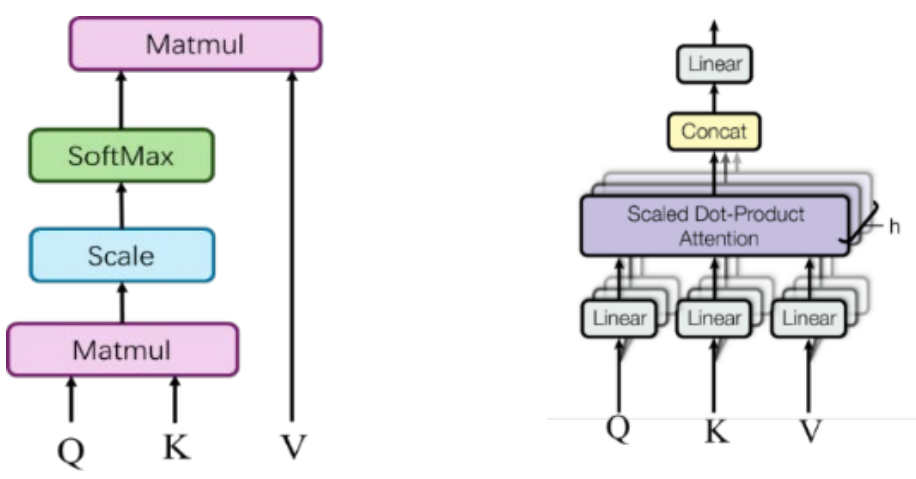

(a) Self-Attention Mechanism (b) Multi-Head Attention Mechanism

Fig. 2. The architecture of the attention mechanism

While the linear transformations are consistent across different positions within the same layer, they utilize different parameters from one layer to another.

### B. Development of LLMs

Building on the transformer architecture, LLMs extend this framework through significant scaling in both model size and training data [10]. While the core architecture, stacked layers of multi-head attention and feedforward networks, remains largely unchanged, several components have been adapted or refined to improve performance and scalability at large scale [11]. Most LLMs are pretrained on diverse corpora, including web text, code, and scientific documents, using self-supervised objectives such as next-token prediction.

At the input level, many models retain the use of token and positional embeddings but introduce refinements. For example, GPT-2 and GPT-3 employ learned positional embeddings [12], while newer models like RoPE (used in LLaMA and ChatGLM) adopt rotary position encoding to improve extrapolation to longer sequences [11], [13], [14]. Tokenization strategies have also evolved: earlier models use byte pair encoding (BPE) [15], but more recent models such as T5 use sentencepiece tokenization [16], and GPT-4 uses byte-level or Unicode-aware variants to better handle multilingual and symbolic input [17]. These changes improve the model's robustness to formatting, symbol-heavy text, and low-resource languages.

At the attention level, although the base formulation remains multi-head scaled dot-product attention, large-scale models have introduced various efficiency and stability enhancements. For instance, GPT-NeoX applies attention weight scaling and recomputation to manage memory [18], while PaLM and BLOOM improve numerical stability through normalization techniques and precision control during training [19], [20]. Some models also explore sparse attention patterns or grouped-query attention (e.g., GQA in Falcon), which reduce computational cost at inference time while maintaining quality [21], [22].

Together, these architectural and technical adaptations permit efficient large-scale operation and confer cross-domain and cross-task generalization capability in the absence of task-specific tuning.

## III. Prompt Framework and Experimental Setups

### A. Prompt Framework Design

To enable LLM-based AD, we design a standardized and modular prompt framework that converts numeric input into structured text suitable for LLMs [23]. It is applied consistently across all experimental settings. The prompt incorporates a rule-based decision process using the three-sigma criterion, which flags sensor values beyond ±3 standard deviations (std) as anomalies. The full structure is outlined in the Prompt Template:

*1) Role Instruction:* Each prompt begins with a role-defining instruction to guide the LLM's behavior. For example: *"You are a power system analyst. Your task is to determine whether the following sample indicates an anomaly."* This framing establishes context for the model and aligns its output with task expectations.

*2) System Context:* The system context provides background information on the input data. We briefly describe the IEEE 14-bus system and the types of features observed: *"The dataset includes real/reactive power injections ($P_i$, $Q_i$), line power flows ($P_{ij}$, $Q_{ij}$), and bus voltage magnitudes (V), with 68 features per sample."* This ensures that the model understands the domain-specific nature of the data.

*3) Anomaly Rule*: The prompt includes an explicit statistical rule for AD, based on the 3-sigma criterion applied to z-score, as shown in the Prompt Template. By embedding the rule into the prompt, we encourage consistent, interpretable reasoning aligned with standard AD practices.

*4) Optional Examples (Few-shot / ICL Only):* In few-shot and ICL settings, we prepend the prompt with $k \in \{2,10\}$ labeled examples. Each example includes an input block, followed by the correct output. These are sampled from the training set and selected to represent both normal and anomalous cases with varying anomaly strengths.

*5) Value Block:* The value block is presented as a plain-text table, where each row corresponds to a single sensor. Each row includes the sensor name, its current value, the mean and std of dataset, and the computed z-score.

*6) Output Schema:* To ensure consistency and interpretability, the model is instructed to output exactly two lines: ***i)*** a one-word label ("normal" or "anomaly") and ***ii)*** a brief explanation for the classification.

### B. Experiment Setup

To assess the anomaly detection capabilities of large language models under various prompting strategies, we design five experimental settings: zero-shot, few-shot, ICL, LoRA fine-tuning, and a hybrid LLM-traditional method.

*Zero-shot*: In the zero-shot setting [24], the model receives only descriptive information for the test sample without any labeled examples. This setting evaluates whether the LLM can follow statistical instructions and deduce the necessary information without prior demonstrations.

We design four variants of Value Block to assess how different levels of numeric information affect model performance. The first variant includes only the rule description and the raw feature values, without mean or std. The second adds the mean and std for each feature. The third further includes the calculated z-scores. In the fourth variant, only z-scores are provided, omitting the raw values, means, and std.

**Prompt Template**

**Role Instruction**: You are a power system analyst. Your task is to determine whether the following sample indicates an anomaly.
**System Context**: The dataset includes real/reactive power injections ($P_i$, $Q_i$), line power flows ($P_{ij}$, $Q_{ij}$), and bus voltage magnitudes ($V$), with 68 features per sample.
**Anomaly Rule**: 3-sigma
- You will receive 68 sensor values.
- If the mean, std, or z-score is not provided, you must infer them from the available values before making a decision.
- Decision procedure:
  - For each measurement $i$, compute z-score = (value - mean) / std, where std := max(std, 1e-12).
  - Label as "anomaly" if at least one measurement satisfies $|z| \geqslant 3.0$; otherwise, label as "normal".
  - The Value Block below includes value, mean, std, and absolute z-score for each measurement, grouped by task category.

** Examples (Optional) **:{
Example1: value_block, label;
Example2: value_block, label;
…}
**Value Block**: {value_block}
**Output Format**: (must be exactly two lines):
1) Label: normal OR anomaly (1 word only)
2) Brief explanation for the classification
**Answer**:

*Few-shot and ICL*: The few-shot setting extends the zero-shot prompt by including two labeled instances from the training data, one normal and one anomalous, enabling the model to observe a minimal form of reasoning prior to prediction. In the ICL setting [26], the prompt is further enriched with five labeled examples comprising both normal and anomalous cases that reflect diverse anomaly types and feature activation patterns, thereby providing the model with a broader reasoning context for downstream classification.

*LoRA Fine-tuning* [27]: Here, we apply parameter-efficient fine-tuning using LoRA on the training data. The model is supervised to generate both the anomaly label and rationale, allowing it to internalize the task beyond prompting alone. We apply adaptation to the attention layer by targeting the $Q$, $K$, and $V$ projections. A lightweight LoRA module is inserted into each of these components to enable efficient fine-tuning.

*Hybrid Method*: The hybrid method combines LLM predictions with a traditional DL detector. The LLM performs feature selection by identifying the most relevant indicators of abnormality, which are then passed to a DL-based detector for final anomaly classification. This approach aims to test whether LLM-based reasoning can enhance the robustness of traditional methods, particularly in borderline or uncertain cases.

### C. Evaluation Metrics

We evaluate model performance using the Precision–Recall evaluation, and confusion matrix is shown in Fig. 3, where a positive denotes an anomaly and a negative represents normal system operation. The matrix comprises true positives (TP), true negatives (TN), false positives (FP), and false negatives (FN), corresponding to correct and incorrect classifications of anomalies and normal instances. From these, we derive four standard evaluation metrics: accuracy (overall classification correctness), recall (proportion of anomalies correctly detected), precision (proportion of predicted anomalies that are correct), and F1-score (harmonic mean of precision and recall). These four metrics are computed as follows:

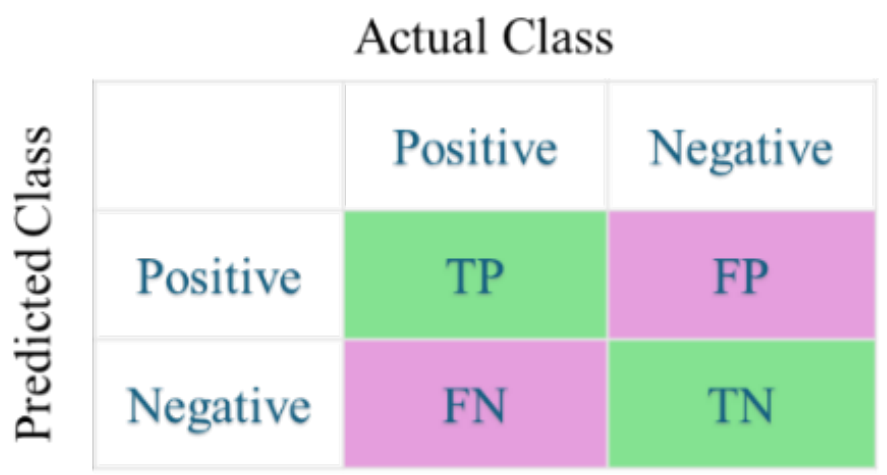

Fig. 3. Confusion matrix

$$Accuracy = \frac{TP + TN}{TN + TP + FN + FP} \tag{9}$$

$$Recall = \frac{TP}{TP + FN} \tag{10}$$

$$Precision = \frac{TP}{TP + FP} \tag{11}$$

$$F1 - score = 2 * \frac{Recall * Precsion}{Recall + Presicion} \tag{12}$$

## IV. Case Study

### A. Simulation System

*1) Dataset:* The proposed method is evaluated on the IEEE 14-bus AC power system. Hourly load profiles from the Electric Reliability Council of Texas (ERCOT) are mapped to the system's load buses [28], [29], and the first 1,000 generated measurements are selected to construct the dataset. Each sample includes 68 features: active and reactive power injections ($P_i$, $Q_i$), line flows ($P_{ij}$, $Q_{ij}$), and voltage magnitudes ($V_i$). Anomalous data are generated by injecting 15% deviations into three randomly selected sensor values, forming compromised states $x_a = x_i + a$. For LoRA fine-tuning, 1,200 samples (600 normal and 600 anomalous) are used. The remaining 400 samples are evenly split into validation and test sets.

*2) LLM model:* In this case study, we adopt GPT-OSS-20B as the representative LLM [30]. GPT-OSS-20B is an open-source, decoder-only Transformer model with approximately 24 billion parameters. It is pretrained using autoregressive language modeling on a large-scale multilingual and domain-diverse corpus. The model is publicly available via the Hugging Face platform and supports efficient adaptation techniques. This choice allows us to evaluate LLM performance on structured numeric AD in power systems under realistic resource constraints. All experiments are conducted on a workstation equipped with two NVIDIA RTX 6000 Ada GPUs, each with 48 GB of memory.

### C. Performance Evaluation

This section presents the evaluation results of LLMs for numeric AD in power systems. We begin with an ablation study on the zero-shot configuration to assess the impact of varying prompt content, as subsequent experiments inherit the most effective prompt structure. We then compare five experimental settings: zero-shot, few-shot, ICL, LoRA fine-tuning, and a hybrid LLM with traditional methods, all under a standardized prompt framework. Finally, we evaluate whether LLM-based reasoning enhances the robustness of traditional DL detectors in the cyber-physical power grid applications.

*1) Ablation Study on Prompt Structure (Zero-Shot Setting):*

Table I reports the zero-shot performance of LLMs under four prompt structures, each varying the Value Block content. Using only raw sensor values yields poor recall (12.0%) and a low F1-score (20.2%), indicating limited generalization without contextual information. Adding the mean and std increases recall to 56.5% but offers limited gains in overall performance. Incorporating z-scores alongside raw values and statistics improves all metrics, particularly recall (79.5%). The best performance is achieved using only z-scores, suggesting that concise, normalized inputs are more effective than verbose statistical descriptions. Overall, the F1-score shows a consistent upward trend as richer statistical context is added, underscoring the importance of prompt design in balancing precision and recall. Based on these findings, the z-score format is used in all subsequent experiments.

Table I. Zero-shot ablation results under different prompt structures

| Value Block Types | Accuracy (%) | Recall (%) | Precision (%) | F1-score (%) |
| --- | --- | --- | --- | --- |
| Value | 52.5 | 12.0 | 63.2 | 20.2 |
| Mean, Std, Value | 52.2 | 56.5 | 52.1 | 54.2 |
| Mean, Std, Value, Z_score | 60.5 | 79.5 | 57.6 | 66.8 |
| Z_score | 78.5 | 64.5 | 89.6 | 75.0 |

*2) Comparison Across Prompting and Adaptation Strategies:*

Table II presents the anomaly detection performance of five LLM-based configurations: zero-shot, few shot, ICL, LoRA fine-tuning, and a hybrid LLM combined with traditional methods.

F1-score increases consistently from 75.0% in the zero-shot setting to 97.2% in the hybrid configuration, highlighting the benefits of progressively incorporating supervision and adaptation. In the zero-shot setup, the model achieves high precision (89.6%) but low recall (64.5%), meaning it raises few false alarms but misses many true anomalies, which is an unacceptable trade-off in power systems. Few-shot prompting raises recall to 88.0%, indicating it detects more anomalies but introducing more false positives. ICL narrows the precision-recall gap and improves overall detection reliability. LoRA fine-tuning boosts recall to 99.0%, nearly eliminating missed detections, but reduces precision to 72.3%, thereby increasing false alarms. The hybrid LLM-traditional model resolves this trade-off, achieving both high recall (96.5%) and precision (98.0%), making it a robust solution for numeric anomaly detection. These findings highlight that modest adaptation can substantially improve LLM performance, especially when combined with traditional detectors.

*3) Comparison Between Traditional and LLM-Enhanced Hybrid Detectors:*

Table III presents a comparison between a standalone DL detector and an LLM-enhanced hybrid model, where the LLM performs rule-based feature filtering before DL classification. The traditional DL model detects most anomalies but suffers from frequent false alarms due to its limited precision (80.3%). With LLM-based reasoning layered on top, the hybrid model achieves a much better balance with an F1-score of 97.2%, which significantly reduces false positives while maintaining high detection coverage. Overall, the hybrid approach offers a clear improvement in both accuracy and reliability.

Table II. Performance of LLM-Based Methods for AD Under Different Prompting Paradigms

| Prompt paradigm | Accuracy (%) | Recall (%) | Precision (%) | F1-score (%) |
|---|---|---|---|---|
| Zero-shot | 78.5 | 64.5 | 89.6 | 75.0 |
| Few-shot | 77.5 | 88.0 | 72.7 | 79.6 |
| ICL | 81.5 | 86.5 | 78.6 | 82.4 |
| Fine-tuned | 80.5 | 99.0 | 72.3 | 83.5 |
| Hybrid LLMs | 97.3 | 96.5 | 98.0 | 97.2 |

Table III. Comparison between traditional methods and LLM-Enhanced Hybrid detectors

| Model | Accuracy (%) | Recall (%) | Precision (%) | F1-score (%) |
|---|---|---|---|---|
| Traditional DL | 87.0 | 98.0 | 80.3 | 88.3 |
| LLM + DL | 97.3 | 96.5 | 98.0 | 97.2 |
| Performance lift | 11.84 | -1.53 | 22.07 | 10.08 |

## V. Conclusions

LLMs' effectiveness on structured numeric telemetry, particularly for AD, remains underexplored in power systems. This study presents a comprehensive evaluation of LLMs for numeric AD using GPT-OSS-20B on the IEEE 14-bus system, covering zero-shot, few-shot, ICL, LoRA fine-tuning, and a hybrid LLM-traditional model.

The results demonstrate that while LLMs benefit from structured numeric cues, concise and well-designed prompts yield stronger generalization, eliminating redundancy can raise the F1-score from 66.8% to 75.0%. Detection performance improves consistently with increasing levels of supervision, confirming that LLMs can effectively adapt to structured numeric tasks with incremental guidance. Furthermore, the hybrid architecture that integrates LLM reasoning with traditional detectors achieves the best trade-off between accuracy and reliability, reaching an F1-score of 97.2% and demonstrating its effectiveness in simulation-based power system scenarios. Future work will extend this framework to larger-scale power systems, incorporate temporal and multi-modal inputs, and assess LLM robustness under dynamic operating conditions.

## Acknowledgment

This material is based upon work supported by the U.S. National Science Foundation under Grant No. 2146156 and No. 2316355. Computing resources for this research were supported by the Office of Naval Research under Grant No. N00014-23-1-2777.